\documentclass[aps,prl,twocolumn,superscriptaddress,twoside,floatfix,amsmath,showpacs,nofootinbib]{revtex4}

\usepackage{graphicx}
\usepackage{amsmath}
\usepackage{textcomp}
\usepackage{dcolumn}
\usepackage{bm}

\usepackage{amssymb}
\usepackage[latin1]{inputenc}
\usepackage{amsfonts}
\usepackage{lineno}
\usepackage{url}
\usepackage{array}
\usepackage{verbatim}
\usepackage{subfigure}
\usepackage{listings}
\usepackage{color}
\usepackage{lscape}
\usepackage{setspace}
\usepackage{textcomp}
\usepackage{footmisc}
\usepackage [english]{babel}
\usepackage [autostyle, english = american]{csquotes}
\MakeOuterQuote{"}

\DeclareMathSizes{10}{9}{6}{6}

\begin{document}


\title{Novel Concept for a Neutron Electric Charge Measurement using a\\ Talbot-Lau Interferometer at a Pulsed Source}

\author{Florian M. Piegsa}
\email{florian.piegsa@lhep.unibe.ch}
\affiliation{Laboratory for High Energy Physics and Albert Einstein Center for Fundamental Physics,\\ University of Bern, CH-3012 Bern, Switzerland }

\date{\today}

\begin{abstract}

A concept to measure the neutron electric charge is presented which employs a precision Talbot-Lau interferometer in a high-intensity pulsed neutron beam.
It is demonstrated that the sensitivity for a neutron charge measurement can be improved by up to two orders of magnitude compared to the current best direct experimental limit.


\end{abstract}












\maketitle

\section{INTRODUCTION}

The quest to determine the electric charge of the neutron is connected to the question of the neutrality of atoms and bulk matter \cite{Unnikrishnan2004}.
In 1959, Lyttleton and Bondi proposed that the expansion of the universe could be accounted for by a slight charge asymmetry between the electron and the proton \cite{Littleton1959}. This hypothesis was quickly disproved by means of a gas efflux experiment using argon and nitrogen \cite{Hillas1959}.
From this and similar indirect measurements a residual neutron charge of less than $10^{-21}$~e can be deduced, where e is the elementary electron charge.
Nevertheless, the questions of charge quantization and of the neutrality of neutrons, neutrinos, and atoms remain under debate \cite{PhysRevD.49.3617,PhysRevD.44.3706,PhysRevD.48.4481,PhysRevD.51.2411,PhysRevLett.100.120407}. 
Direct measurements of the neutron charge are additionally motivated by the possibility that the charge of a free particle might be slightly different in magnitude compared to its charge when bound in an atom \cite{PhysRev.153.1415}.
One immediate consequence of a finite neutron charge would be that a speculative neutron to anti-neutron oscillation is forbidden if charge conservation is valid \cite{Baldo1994,Frost2016}.
Direct measurements have been performed using cold neutrons passing through a strong electric field oriented perpendicular to the beam direction \cite{PhysRev.153.1415,PhysRevD.25.2887}. In such an experiment, one would expect a corresponding transverse beam deflection if a hypothetical neutron charge is present. This method was brought to perfection in an heroic experiment by Baumann and colleagues leading to the present best direct limit on the neutron charge of  
$(-0.4 \pm 1.1) \times 10^{-21}$~e \cite{Baumann1988}. Measurements using the same technique adapted to UCN have been performed, but were not yet able to compete with the sensitivity of the aforementioned experiment due to technical complications \cite{Borisov1988,Plonka2010,PhysRevD.97.052004}. 
Presently, there are prospects for new neutron electric charge efforts with the ultimate goal to improve the current limit by more than one order of magnitude. 
One proposal plans to employ the technique of precision UCN gravity resonance spectroscopy \cite{Jenke2011,Jenke2014,Jenke201467}. Here, the energy eigenstates of neutrons in the gravity potential of the Earth are probed by bouncing them on a horizontal mirror with a simultaneously applied electric field \cite{Durstberger2011}.
Another idea suggests to measure the deflections of a cold neutron beam by means of the spin-echo small angle neutron scattering technique \cite{Voronin2013}. \\
In 1962, Maier-Leibnitz and Springer performed the first experiments using a neutron Fresnel bi-prism interferometer \cite{MaierLeibnitz1962}. In their publication, they already proposed many possible applications of such a new device, among which was also the measurement of the neutron charge and the investigation of the gravitational interaction of the neutron.
Later, the method of Mach-Zehnder type neutron interferometers made from silicon single crystals was pioneered by Rauch, Treimer, and Bonse \cite{RAUCH1974369,Rauch2015}. Neutron interferometry is a powerful tool with many applications in a multitude of research topics. They range from direct tests of quantum mechanics to searches for exotic interactions, precision measurements of neutron scattering lengths, investigation of topological phases, and observation of effects due to the Earth's gravitational field on neutrons \cite{PhysRevLett.34.1472,PhysRevA.56.1767,Ioffe1998si,Hasegawa2001,Hasegawa2003,Bartosik2009,PhysRevLett.102.200401,Sponar2010,Denkmayr2014,Lemmel2015310,1367-2630-17-2-023065,Denkmayr2017}.
Single crystal interferometers have the advantage of intrinsic perfect alignment. On the other hand their field of application is limited to short centimeter-size setups and to thermal neutron energies due to the necessary Bragg diffraction process. 
Some applications, like the measurement of the neutron charge, however, would benefit from longer interaction times and thus from meter-size setups. 
Another option to increase the interaction time of the neutrons and hence the sensitivity is to use phase grating interferometers for very cold neutrons \cite{Gruber1989,Eder1989,vanderZouw2000}. However, they have the drawback of lower neutron flux and that very cold neutrons already fall appreciably in the Earth's gravitational field, which limits again the length of such a setup. \\
Recently, neutron absorption and phase gratings on silicon or quartz substrates with pitches on the order of micrometers have been developed \cite{Grunzweig2008,Grunzweig2007a}. They are commonly used in neutron phase contrast radiography and neutron dark-field imaging in various physics and industry applications \cite{Grunzweig2006,Strobl2008,Grunzweig2013,Betz2016,Harti2017}. 
Here, it is proposes to employ such gratings for the application in a novel so-called Talbot-Lau interferometer which consists of three well-aligned neutron absorption gratings placed at certain distances to each other \cite{Lau1948}. 
This type of interferometer will represent an entirely new precision instrument in the field of neutron physics.\footnote{Recently, a related so-called far-field Moir\'e neutron interferometer using phase gratings has been realized for the first time \cite{PhysRevLett.120.113201}. In this first proof-of-principle study it was demonstrated that a meter-long setup can be actually achieved. 
Another similar instrument, namely a Moir\'{e} deflectometer, is used in the AEgIS anti-hydrogen gravity experiment \cite{Aghion2014,PhysRevA.54.3165}.}
It will be an ideal tool to detect tiny deflections of a neutron beam applicable in a variety of experiments, e.g.\ searching for a non-zero neutron electric charge, performing ultra small angle scattering measurements, and investigating the deflection of neutrons due to gravitational interaction with macroscopic test masses \cite{Frank2009,Abele2012}.
Theoretical calculations and simulations of such kind of grating interferometers are described in detail in \cite{PhysRevA.78.013601,PhysRevA.94.033625}. 
Here, the concept of a meter-long symmetric neutron Talbot-Lau interferometer operated in time-of-flight mode is presented. 
Ultimately, such an interferometer can be used to perform a future high-precision measurement of the neutron electric charge at the proposed high-intensity fundamental physics beam line ANNI at the European Spallation Source  \cite{Klinkby2016}.






\section{MEASUREMENT PRINCIPLE}

\begin{figure}
	\centering
		\includegraphics[width=0.480\textwidth]{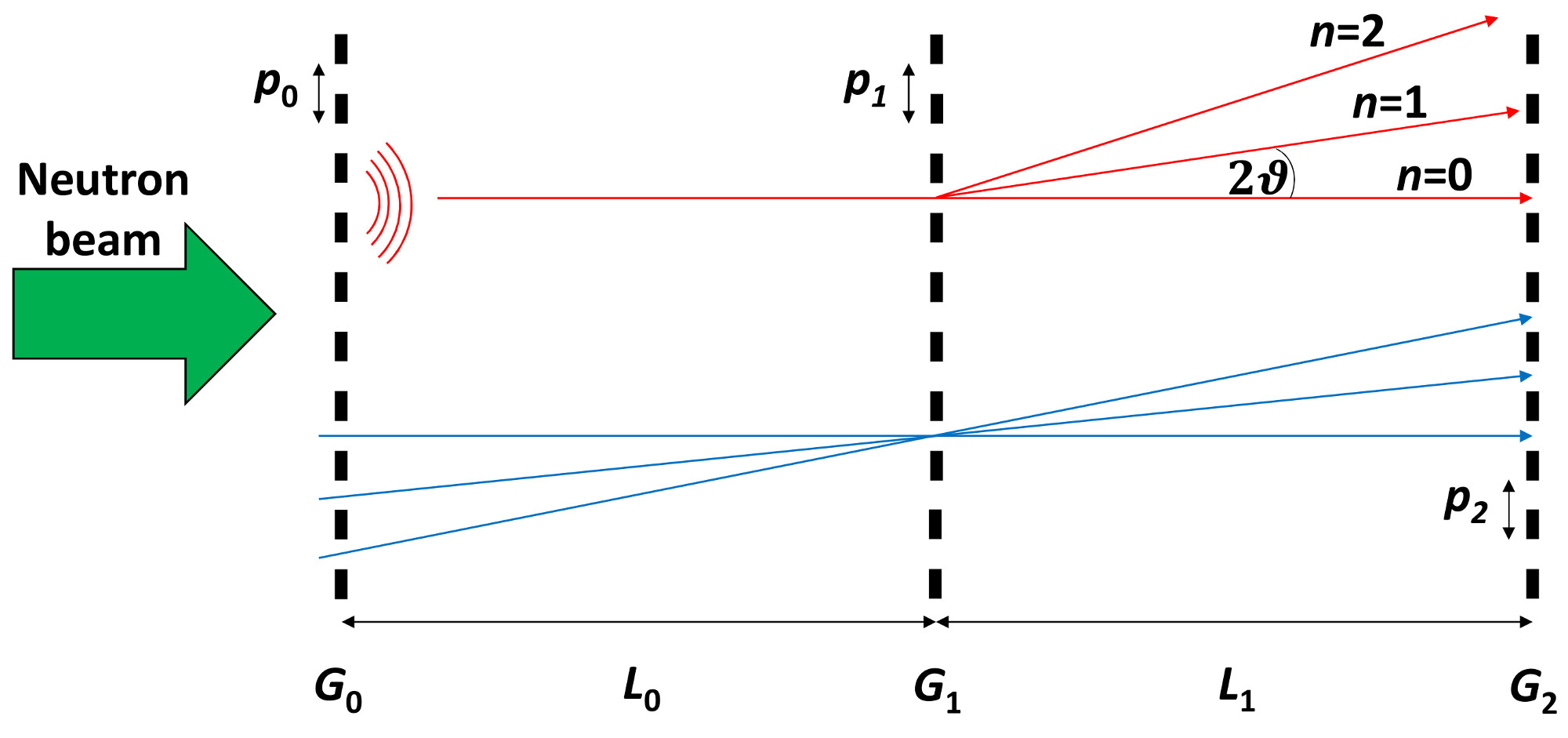}
	\caption{(Color online) Scheme of a neutron Talbot-Lau interferometer with three absorption gratings with periods $p_0$, $p_1$, and $p_2$ and distances $L_0$ and $L_1$. The grating $G_0$ produces coherent line sources from the incoming incoherent and divergent neutron beam. The red arrows (above) indicate neutron Bragg diffraction at the second grating with diffraction orders $n$ and the corresponding diffraction angle $2 \vartheta$. 
	The blue arrows (below) indicate ballistic zero-order trajectories. In a symmetric configuration the setup consists of three identical gratings with $p_0=p_1=p_2$ and $L_0=L_1$.}
	\label{fig:TL-Interferometer}
\end{figure}

In Fig.\ \ref{fig:TL-Interferometer}, a schematic drawing of a Talbot-Lau interferometer adapted to neutrons is presented.
It consists of three neutron absorption gratings $G_0$, $G_1$, and $G_2$ with their associated periods $p_i$. 
The duty cycles of the gratings, which describe the ratio of the width of the slit openings and the periods, are given by their values $R_i$.
In the following, one considers the special case of a symmetric setup where all three gratings are identical and have an equal period $p$ as well as a fixed duty cycle $R$, and are placed at a distance $L$ to each other. \\
Before the performance of the interferometer is discussed, one needs to consider the definitions of transverse and longitudinal (or temporal) coherence \cite{vanderVeen2004}.
The purpose of the grating $G_0$ is to convert an initially incoherent beam of neutrons into multiple coherent line sources.
The resulting transverse coherence length at the position of the grating $G_1$ is given by
\begin{equation}
  \xi_t = \frac{L \lambda}{R p}
\label{eq:transv-coherence}
\end{equation}
where $\lambda$ is the neutron de Broglie wavelength. On the other hand, the longitudinal coherence of the beam depends on the actual width $\Delta \lambda$ of the employed wavelength spectrum
\begin{equation}
  \xi_l =\frac{\lambda^2}{2 \Delta \lambda} 
\label{eq:long-coherence}
\end{equation}
The beam is diffracted by the grating $G_1$ according to Bragg's law such that at the position of the grating $G_2$ an interference pattern arises. 
Additionally, each individual coherent line source of $G_0$ contributes to the intensity of this combined pattern and the interferometer allows for employing divergent beams depicted as ballistic trajectories in Fig.\ \ref{fig:TL-Interferometer}. 
The periodicity of the interference fringes is of the same size as the grating period $p$ and usually cannot be resolved with a standard position sensitive neutron detector or neutron CCD camera. Instead one employs a third so-called analyzer grating $G_2$ and an integral neutron detector placed behind it. In this measurement scheme the grating $G_2$ is scanned in small steps perpendicular to the neutron beam direction and the oscillating intensity pattern is recorded. Any deflection of the neutron beam between $G_1$ and $G_2$ induces a corresponding shift of the interference pattern and can thus be detected. \\
Another important characteristic quantity is the so-called Talbot length $L_{\text{T}} = p^2/\lambda$, which describes the distance at which the gratings need to be placed to cause that the first order Bragg interference maximum of the wavelength $\lambda$ is diffracted exactly by one period. 
In order to avoid that these Bragg peaks potentially wash out the interference pattern the grating distance $L$ has to be a multiple $m$ of the Talbot length, i.e.\ $L=m \cdot L_{\text{T}}$ with $m \in \{1,2,3,...\}$. As a consequence, given a fixed distance between the gratings the interferometer fulfills this condition only at specific neutron wavelengths $\lambda = m p^2 / L$. From this follows a transverse coherence of $ \xi_t = m p / R$.
\begin{figure}
	\centering
		\includegraphics[width=0.48\textwidth]{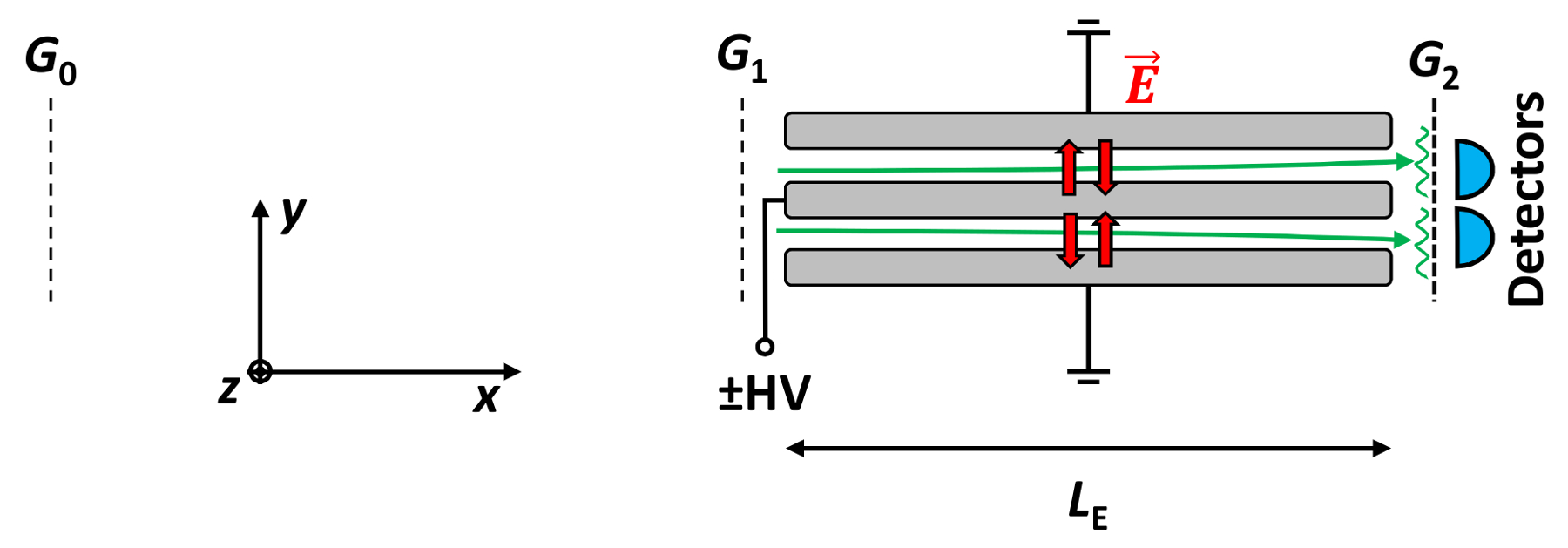}
	\caption{(Color online) Top-view scheme of a Talbot-Lau interferometer setup to measure the neutron electric charge. Three electrodes, one on high-voltage and two connected to ground potential, are located between gratings $G_1$ and $G_2$. The two beams between the electrodes are experiencing opposite electric field directions and are detected with two separate detectors. Hence, in case of a non-zero neutron electric charge their beam paths are bent in opposite directions along the $y$-axis. }
	\label{fig:InterferoSetup}
\end{figure}
Moreover, to avoid that the second order Bragg peaks cause a decrease in signal visibility the width of the wavelength spectrum needs to be limited to $\Delta \lambda < \frac{p^2}{2L}$ 
or $\xi_l > m\lambda$, respectively.\footnote{Recently, neutron ultra small angle scattering measurements have been performed on an absorption grating with a period of 2~\textmu m. These tests showed that the relative peak intensity of even higher order diffraction maxima is below 1\% and can thus be neglected.}
These constraints essentially limit the usable neutron wavelength to specific narrow bands and thus also the overall intensity of the neutron beam.
However, in the novel concept presented this drawback can be overcome by operating the interferometer in time-of-flight mode and at a pulsed spallation source.  
Hence, one can regain intensity and profit from a broad wavelength distribution by time-tagging the neutrons' arrival/flight times.
The time-of-flight method also provides the means to investigate possible velocity dependent systematic effects.\\
The method of measuring an electric charge of the neutron $Q$ with the interferometer is straight forward. Applying an electric field $E$ with alternating high-voltage polarity transversely to the neutron flight path causes a deflection of the beam and equally a shift of the interference pattern by 
\begin{equation}
 \Delta y = \frac{m_{\text{n}} Q E L^2 \lambda^2  }{(2 \pi \hbar)^2} 
\label{eq:chargeshift}
\end{equation}
where $m_{\text{n}}$ is the neutron mass and $\hbar$ is the reduced Planck constant.
Here, it is assumed that the length $L_{\text{E}}$ of the electrodes producing the field is approximately equal to the distance between the gratings $L$.\footnote{The orientation of the electric field should be horizontal to avoid potential systematic effects due to the Earth's gravitational field.}
For instance, assuming a field strength $E=100$~kV/cm with an electrode gap width $d=1$~cm, a length $L=5$~m, a fixed neutron wavelength $\lambda=0.5$~nm, and a charge $Q = 10^{-23}$~e (two orders of magnitude smaller than the current experimental limit) yields a deflection of 0.5~pm between the two electric field polarity states.\footnote{Note, in comparable experiments with similar configurations fields of $80-130$~kV/cm and 60~kV/cm were reached, respectively \cite{Dress1977,Baumann1988}.}
In Fig.\ \ref{fig:InterferoSetup} a sketch of the proposed experimental setup is presented. It employs three electrodes placed between $G_1$ and $G_2$ and two separate neutron beams which experience electric fields with anti-parallel orientations. In case of a non-zero neutron electric charge this causes deflections of the beams in opposite directions along the $y$-axis. Hence, the use of two beams allows to correct and compensate for potential common-mode drifts and noise. 
Furthermore, the two beam method provides an ideal way to normalize the count rates in order to compensate for potential variations in the neutron source flux.

\section{STATISTICAL SENSITIVTIY}

\begin{figure}
	\centering
		\includegraphics[width=0.450\textwidth]{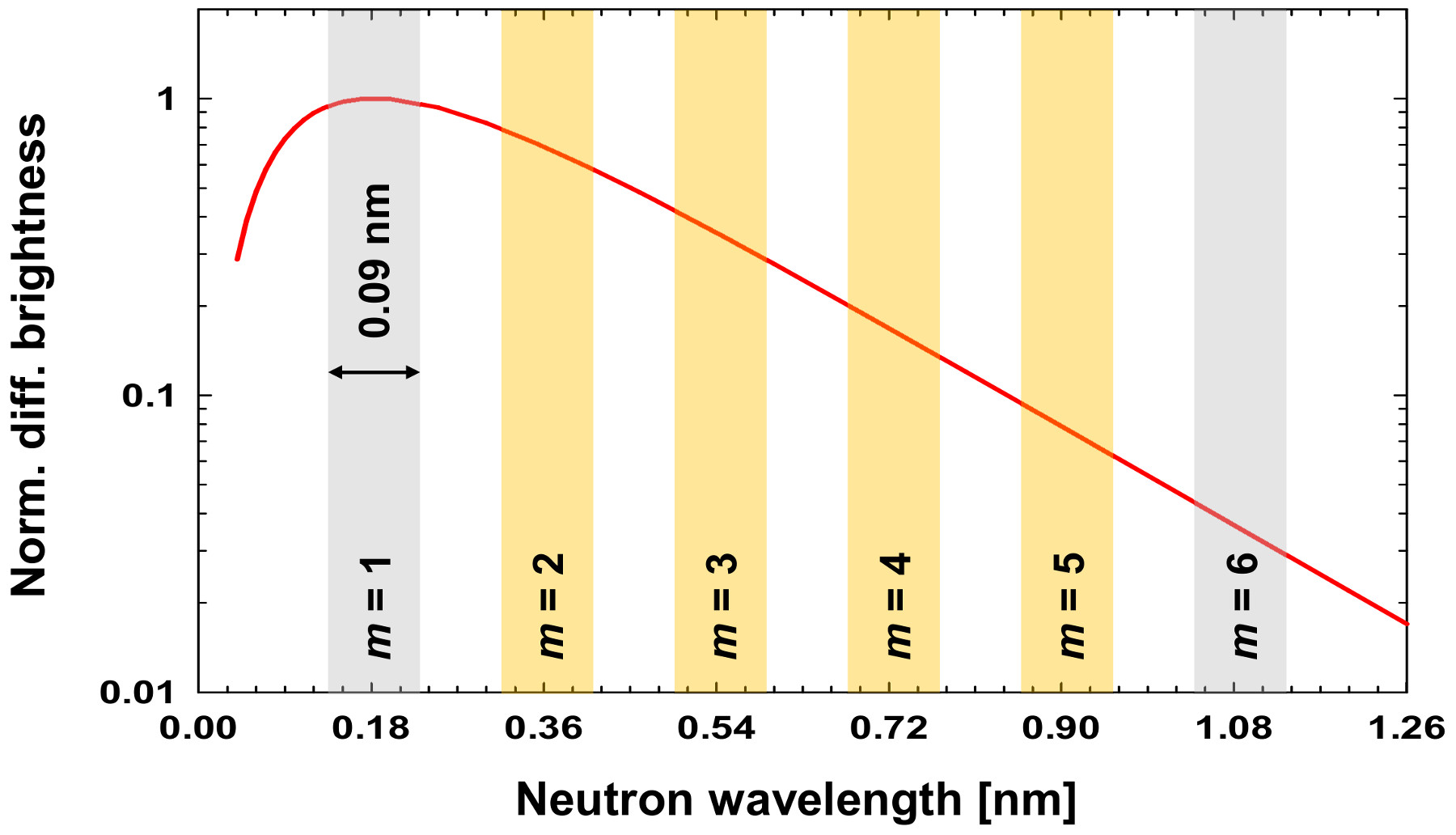}
	\caption{(Color online) Normalized differential neutron particle brightness as a function of $\lambda$ at the PF1b beam line at ILL, adapted from \cite{Abele2006}. The red curve represents the fit curve to the measured data. The marked regions indicate the 0.09~nm wide wavelength bands as discussed in the text. The light yellow regions correspond to $m=2,3,4,$ and $5$ which are the bands where pulse frame overlap will be suppressed at the proposed ANNI beam line. The small ticks on the horizontal axis are separated by the the mentioned time-of-flight wavelength resolution of 0.03~nm. }
	\label{fig:lambda}
\end{figure}

The statistical sensitivity (standard deviation) of a Talbot-Lau interferometer for detecting a beam deflection is independent of the neutron wavelength and is given by
\begin{equation}
  \sigma (\Delta y) = \frac{p}{\eta \pi \sqrt{N}}
\label{eq:sensi-shift}
\end{equation}
where $\eta$ is the interference fringe visibility and $N$ is the total number of detected neutrons. 
With Eq.\ (\ref{eq:chargeshift}) this yields a statistical sensitivity on the neutron charge of
\begin{equation}
  \sigma (Q) = \frac{4 \pi \hbar^2 p}{\eta m_{\text{n}} E L^2 \lambda^2 \sqrt{N}} 
\label{eq:sensi-Q}
\end{equation}
Let us now consider a Talbot-Lau interferometer intended for the European Spallation Source with a length $L=5$~m, a period $p=30$~\textmu m, and a duty cycle $R=50$\%.
This yields equidistant wavelength bands centered at $\lambda = m \cdot 0.18$~nm and a width $\Delta \lambda < 0.09$~nm.
Hence, at the proposed high-intensity fundamental physics beam line ANNI with a broad cold neutron spectrum, one can measure at $m$ from 2 up to 5 without pulse frame overlap and one can achieve a sufficient wavelength resolution of better than 0.03~nm.
The latter assumes a source pulse frequency of 14~Hz and a pulse length of 3~ms with a total source-to-detector distance of 45~m.
Presuming ANNI will have a comparable time integrated neutron flux as the fundamental physics beam line PF1b at ILL and a beam cross section of $10$~cm$^2$ for each partial beam, unpolarized neutron rates behind the interferometer between $10$ and $100$~MHz can be deduced from the data presented in Ref.\ \cite{Abele2006}. With a visibility $\eta = 0.75$ this corresponds to a statistical sensitivity on the beam deflection of 1 to 4~nm in one second or 4 to 15~pm in one day of measurement time, respectively. 
As presented in Fig.\ \ref{fig:lambda}, the neutron beam brightness typically decreases toward longer wavelengths, however, this reduction in intensity is almost entirely compensated by the gain in sensitivity which scales with the $\lambda^{-2}$, compare Eq.\ (\ref{eq:sensi-Q}). 
With this statistical sensitivity and assuming an electric field reversal of typically once every 100~s, thus, requires a stability of the interferometer and the interference fringes of better than 0.1~nm on the same time scale. Employing the two beam method relaxes the situation significantly, since only relative deflections of the beams are considered. Finally, assuming 100~days of data taking, leads to a statistical sensitivity of about 0.5~pm and a corresponding sensitivity on the neutron electric charge of $10^{-23}$~e, respectively.

\section{SYSTEMATIC EFFECTS}

The key factors of such a high-precision measurement are the stability of the instrumental components and the control of their susceptibility to temperature and external vibrations. 
In principle, this also includes the position reproducibility of the scanning mechanism of the gratings. However, instead of performing an entire scan of their relative position, the gratings can be permanently fixed in the position in which the interferometer has its best sensitivity, i.e.\ at the steepest slope of the interferometric pattern, similar to the method applied in neutron electric dipole measurements \cite{Golub1991}.
To keep the entire Talbot-Lau interferometer as stable as possible it should be housed in a vacuum chamber with a temperature stabilized thermal shield. In addition, the vacuum serves the purpose to minimize the scattering in air and the associated reduction of signal visibility.
For the accurate adjustment and alignment of the gratings, one should employ piezo stepping stages with a minimal incremental motion in the 100~nm range, i.e.\ much smaller than the grating period. The angular alignment precision of the gratings needs to be on the order 0.1~mrad or better, assuming a period $p=30$~\textmu m and a slit length of 10~cm. The latter can be adjusted and even be monitored by observing the optical diffraction pattern of a laser beam passing through the individual gratings. For this method it is advantageous to use quartz wafers instead of silicon wafers, since they are transparent for light and neutrons. Furthermore, the entire apparatus should be mounted on a vibration-damped table. \\
One systematic effect already described in the publication by Baumann et al.\ arises from the magnetic interaction of the neutron magnetic moment with a gradient of an electric field in $z$-direction via the relativistic $v \times E$-effect \cite{Baumann1988}. 
Such an electric field is for instance caused by a small tilt of the electrode plates with respect to each other. An upper limit for the tilting angle $\alpha$ for a fully polarized beam can be estimated by comparing the forces acting on the neutron charge and the neutron magnetic moment. This yields the following condition
\begin{equation}
  \alpha < \frac{Q c^2 d \lambda m_{\text{n}}}{\pi \gamma_{\text{n}} \hbar^2 }
\label{eq:misalignment}
\end{equation}
where $\gamma_{\text{n}}$ is the gyromagnetic ratio of the neutron and $c$ is the speed of light in vacuum.
With $Q=10^{-23}$~e, $d=1$~cm, and $\lambda=0.3$~nm this results in a maximum tilting angle between the electrode plates of 0.1~mrad.
However, this condition can be largely relaxed assuming an unpolarized beam.\\
Another effect could occur due to electrostatic forces acting on the interferometer gratings, since a systematic movement of the gratings coupled to the direction of the electric fields could cause a false neutron charge signal. However, this effect is strongly suppressed firstly by employing the two beam method and secondly due to the symmetric instrument setup consisting of three electrodes.\\
As a side remark, one can employ neutron prisms as test samples to calibrate the Talbot-Lau setup as they cause a small known wavelength dependent refraction of the beam. For instance an aluminum wedge with an angle of 10 degrees causes a deflection of 1~\textmu rad (6~\textmu rad) for neutrons with a wavelength of 0.4~nm (1.0~nm).

\section{CONCLUSION}

A new concept to measure the neutron electric charge using a Talbot-Lau neutron grating interferometer has been presented. 
The interferometric device represents a unique tool in the field of neutron research and can be used to detect the smallest beam deflections.
Similar to the recently described new neutron beam electric dipole moment search, the technique can greatly benefit from the intrinsic velocity information of a pulsed source \cite{Piegsa2013a}.
A full-scale instrument optimized for the beam specifications of the European Spallation Source can lead to a statistical improvement of the neutron electric charge sensitivity by up to two orders of magnitude compared to the present best limit in 100 days of data taking.

\section{ACKNOWLEGMENTS}
The author gratefully acknowledges many fruitful discussions
with C.\ Gr\"unzweig, R.\ Harti, M.\ Jentschel, G.\ Pignol, C.\ Pistillo, and A.\ Soter.  
This work was supported via the European Research Council under the ERC Grant Agreement no.\ 715031 and via the Swiss National Science Foundation under the grant no.\ PP00P2-163663. 




\appendix
\bibliography{piegsa-bibfile}
\end{document}